\documentclass[manuscript,screen,review=false]{acmart}
\AtBeginDocument{%
  \providecommand\BibTeX{{%
    \normalfont B\kern-0.5em{\scshape i\kern-0.25em b}\kern-0.8em\TeX}}}

 \setcopyright{none}
\setcopyright{acmcopyright}
\copyrightyear{2023}
\acmYear{2023}

\acmConference[]{}{}{}
\acmBooktitle{
  }
\acmPrice{}

\usepackage{xspace}

\newcommand{\ie}{\emph{i.e.{\xspace}}}
\newcommand{\eg}{\emph{e.g.{\xspace}}}

\newif\ifshowcomments
\showcommentstrue

\ifshowcomments
\newcommand{\TODO}[1]{{\color{red}{[TODO: #1]}}}
\newcommand{\fel}[1]{{\color{blue}{[Fel: #1]}}}

\newcommand{\kat}[1]{{\color[rgb]{0.3,0.85,0.3}{[Kat: #1]}}}
\newcommand{\zhao}[1]{{\color{orange}{SD: #1}}}
\else
\newcommand{\TODO}[1]{}
\newcommand{\fel}[1]{}
\newcommand{\kat}[1]{}
\newcommand{\zhao}[1]{}
\fi

\begin{document}

\title{Heads-Up Computing: Moving Beyond the Device-Centered Paradigm}




 \author{Zhao Shengdong}
\email{zhaosd@comp.nus.edu.sg}
\affiliation{
   \institution{NUS-HCI Lab, National University of Singapore}
   \city{Singapore}
   \country{Singapore}
   \postcode{117602}
 }

 \author{Felicia Tan}
 \email{felicia_tan@zoho.com}
 \affiliation{
   \institution{NUS-HCI Lab, National University of Singapore}
   \city{Singapore}
   \country{Singapore}
   \postcode{117602}
 }

 \author{Katherine Fennedy}
 \email{katherine.fennedy@gmail.com}
 \affiliation{
   \institution{NUS-HCI Lab, National University of Singapore}
   \city{Singapore}
   \country{Singapore}
   \postcode{117602}
 }


\begin{abstract}
This article introduces our vision for a new interaction paradigm: “Heads-Up Computing”, a concept involving the provision of seamless computing support for human’s daily activities. Its synergistic and user-centric approach frees humans from common constraints caused by existing interactions (e.g. smartphone zombies), made possible by matching input and output channels between the device and human. Wearable embodiments include a head- and hand-piece device which enable multimodal interactions and complementary motor movements. While flavors of this vision have been proposed in many research fields and in broader visions like UbiComp, Heads-Up Computing offers a holistic vision focused on the scope of the immediate perceptual space that matters most to users, and establishes design constraints and principles to facilitate the innovation process. We illustrate a day in the life with Heads-Up to inspire future applications and services that can significantly impact the way we live, learn, work, and play.
\end{abstract}



\begin{teaserfigure}
  \includegraphics[width=\textwidth]{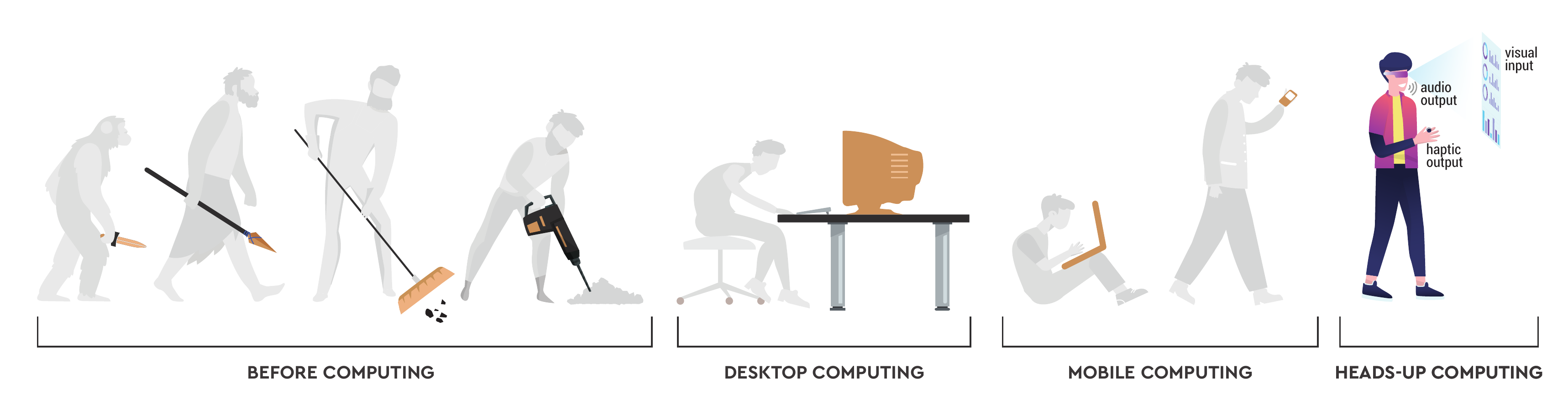}
  \caption{Human's co-evolution with tools.}
  \Description{}
  \label{fig:evolution}
\end{teaserfigure}

\maketitle


Humans have come a long way in our co-evolution with tools.
Well-designed tools effectively expand our physical as well as mental capabilities \cite{CoEvolution}, and the rise of computers in our recent history has opened up possibilities like never before.
The Graphical User Interface (GUI) of the 1970s revolutionized desktop computing; traditional computers with text-based command-line interfaces evolved into an integrated everyday device, the personal computer (PC).
Similarly, the mobile interaction paradigm introduced in the 1990s transformed how information can be accessed anytime and anywhere with a single hand-held device.
Never before have we had so much computing power in the palm of our hands.

The question “Do our tools really complement us, or are we adjusting our natural behavior in order to accommodate our tools?” highlights a key design challenge associated with digital interaction paradigms. 
For example, we accommodate desktop computers by physically constraining ourselves to the desk.
This has, amongst other undesirable consequences, encouraged sedentary lifestyles \cite{Parry2013}and poor eyesight \cite{Sheppard2018}.
While smartphones do not limit mobility, they encourage users to adopt unnatural behavior such as the head-down posture \cite{RegianiBueno2019}.
Users look down at their hand-held devices and pay little attention to their immediate environment.
This `smartphone zombie' phenomenon has unfortunately led to an alarming rise in pedestrian accidents \cite{Zhuang2020}.

\begin{figure}
    \centering
	\includegraphics[width=1.0\textwidth]{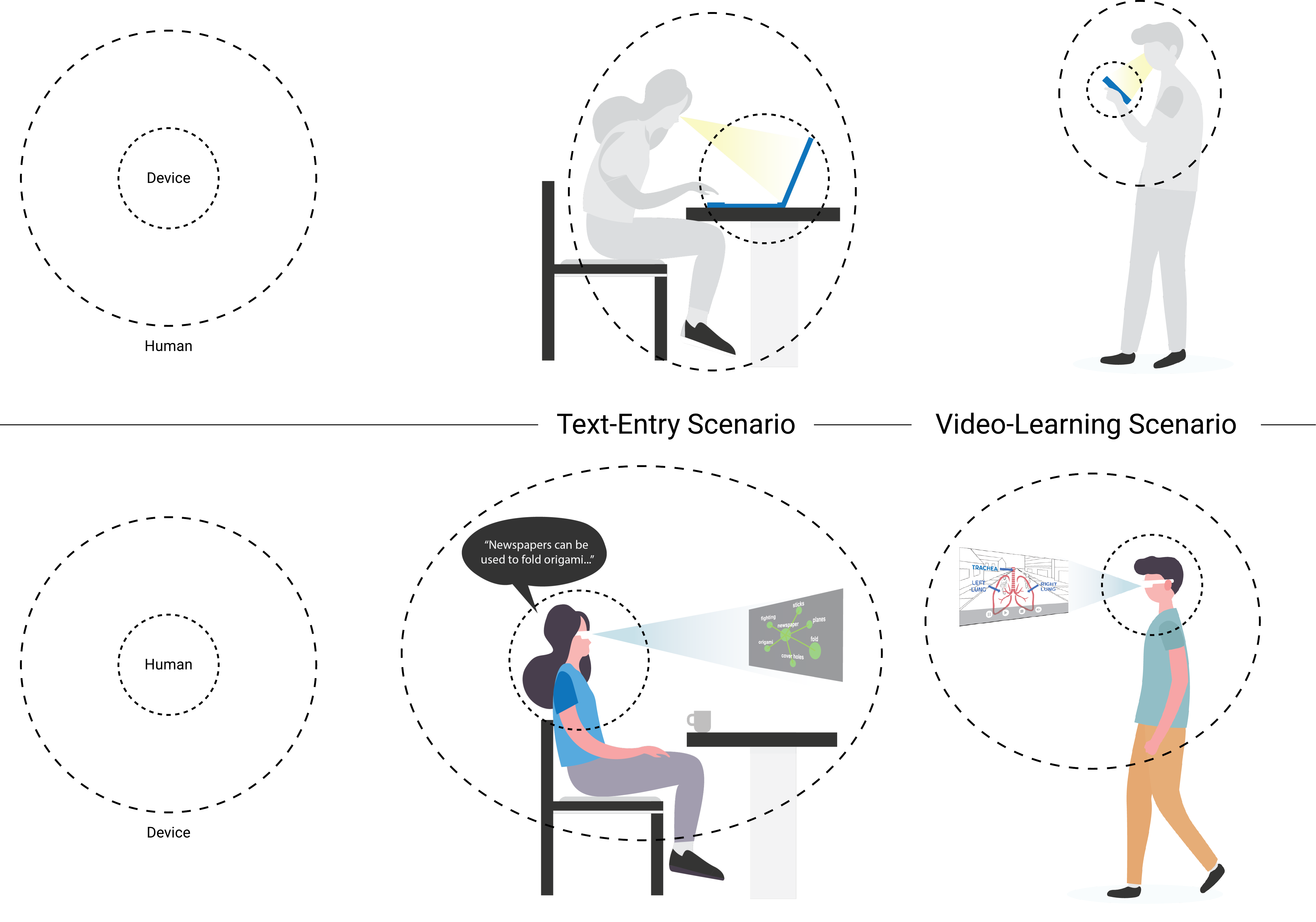}
	\caption{Device-centered (top) vs. human-centered (bottom) interaction in text-entry and video-learning scenarios. We envision that interactivity with digital content can be facilitated by prioritising user contexts (\eg{} walking) and leveraging resources that remain underutilised in these contexts (\eg{} voice)}
	\Description{}
	\label{fig:humanVSdevice} 
\end{figure}

Although there are obvious advantages to the consolidated smartphone hardware form, this `centralization' also means that users receive all inputs and outputs (visual display, sound production, haptic vibration) from a single physical point (the phone).
In addition, smartphones keep our hands busy; users interact with the device by holding it and performing gestures such as typing, tapping, and swiping.
The eyes- and hands-busy nature of mobile interactions limits how users can engage with other activities, and has been shown to be intrusive, uncomfortable, and disruptive \cite{Sapkota2021}.
Could we re-design computing devices to more seamlessly support our daily activities?
Above all, can we move beyond the device-centered paradigm and into a more human-centered vision, where tools can better complement natural human capabilities, instead of the other way around (see Fig.~\ref{fig:humanVSdevice})?

While user-centered design \cite{Norman1986UserCentered} has been introduced for decades, our observations are that everyday human-computer interactions have not aligned with this approach and its goals.
In the next sections, we discuss our understanding of what placing humans at the center stage entails, related works, and how Heads-Up computing may inspire future applications and services that can significantly impact how we live, learn, work, and play.
\section{Humans at the Center Stage}

    \subsection{Understanding the human body and activities}
    
    The human body comprises both input and output (I/O) channels to perceive the world around us.
    Common human-computer interactions involve the use of our hands to click on a mouse or tap a phone screen, or our eyes to read from a computer screen.
    However, the hands and eyes are also essential for performing daily activities such as cooking and exercising.
    When device interactions are performed simultaneously with these primary tasks, competition for I/O resources is introduced \cite{Oulasvirta2005Resources}. 
    As a result, current computing activities are performed either separately from our daily activities (\eg{} work in an office, live elsewhere) or in an awkward combination (\eg{} typing and walking like a smartphone zombie).
    While effective support of multitasking is a complex topic, and in many cases, not possible, computing activities can still be more seamlessly integrated with our daily lives if the tools are designed using a human-centric approach.
    By carefully considering resource availability, \ie{} the amount of resources available for each I/O channel in the context of the user’s environment and activity, devices could better distribute task loads by leveraging underutilized natural resources and lessening the load on overutilized modalities.
    This is especially true for scenarios involving so-called multi-channel multitasking \cite{Eyal2019}: in which one of the task is largely automatic (\eg{} routine manual tasks such as walking, washing dishes, etc.).  
    
    To design for realistic scenarios, we look into \citet{Dollar2014}’s categorization of Activities of Daily Living (ADL), which provides a taxonomy of crucial daily tasks (albeit originally created for older adults and rehabilitating patients).
    The ADL categories range from domestic (\eg{} office presentation), extradomestic (\eg{} shopping), and physical self-maintenance (\eg{} eating), providing sufficient representation of what the general population engages in every day.
    It is helpful to select examples from this broad range of activities when learning about resource demands.
    We can analyze an example activity for its hands- and eyes-busy nature, identify underutilized/overutilized resources, then select opportunistic moments for the system to interact with the user.
    For example, where the primary activity requires the use of hands but not the mouth and ears (\eg{} when a person is doing laundry), it may be more appropriate for the computing system to prompt the user to reply to a chat message via voice instead of thumb-typing.
    But if the secondary task requires a significant mental load, \eg{} composing a project report, the availability of alternative resources may not be sufficient to support multitasking.
    Thus, it is important to identify secondary tasks that not only can be facilitated by underutilised resources, but also minimal overall cognitive load that are complementary to the primary tasks.
    
    To effectively manage resources for activities of daily living and digital interactions, we refer to the theory of multitasking.
    According to \citet{Salvucci2009UnifiedTheory}, multitasking involves several core components (ACT-R cognitive architecture, threaded cognition theory, and memory-for-goals theory).
    Multiple tasks may appear \textit{concurrently} or \textit{sequentially}, depending on the amount of time a person spends on one task before switching to another.
    For concurrent multitasking, tasks are harder to perform when they require the same resources.
    They are easier to implement if multiple resource types are available \cite{Wickens2002}.
    In the case of \textit{sequential} multitasking, users switch back and forth between the primary and secondary tasks over a longer period of time (minutes to hours).
    Reducing switching costs and facilitating the rehearsal of the `problem representation'~\cite{Anderson1998ACTR} can significantly improve multitasking performance.
    Heads-Up computing is explicitly designed to take advantage of these theoretical insights: 1) its voice + subtle gesture interaction method relies on available resources during daily activities; 2) its heads-up optical head-mounted see-through display (OHMD) also facilitates quicker visual attention switches. 
    
    Overall, we envision a more seamless integration of devices into human life by first considering the human's resource availability, primary/ secondary task requirements, then resource allocation.

    \subsection{Existing traces of human-centered approach}
    
    Existing designs like the Heads-Up Display (HUD) and smart glasses exemplify the growing interest in human-centered innovation.
    A HUD is any transparent display that can present information without requiring the operator to look away from their main field of view \cite{Ward1994}.
    HUD in the form of a windshield display has become increasingly popular in the automotive industry \cite{Betancur2018}.
    Studies have shown that it can reduce reaction times to safety warnings and minimize the time drivers spend looking away from the road \cite{Doshi2009}.
    This application ensures the safety of vehicle operators.
    On the other hand, smart glasses can be seen as a wearable HUD that offers additional hands-free capabilities through voice commands.
    The wearer does not have to adjust their natural posture to the device; instead, a layer of digital information is superimposed upon the wearer’s vision via the glasses.
    While these are promising ideas, their current usage is focused more on resolving particular problems and is not designed to be integrated into human’s general and daily activities.
    
    Beyond devices and on the other end of the spectrum, we find general-purpose paradigms like Ubiquitous Computing (UbiComp), which also paints a similar human-centered philosophy but involves a very broad design space.
    Conceptualized by \citet{Weiser1991}, UbiComp aims to transform physical spaces into computationally active, context-aware, and intelligent environments via distributed systems.
    Designing within the UbiComp paradigm has led to the rise of tangible and embodied interaction \cite{Rogers2006}, which focuses on the implications and new possibilities for interacting with computational objects within the physical world \cite{Hornecker2011}.
    These approaches understand that technology should not overburden human activities and that computer systems should be designed to detect and adapt to changes in human behavior that naturally occur.
    However, the wide range of devices, scenarios, and physical spaces (\eg{} ATM spaces) means that there is much freedom to create all kinds of design solutions.
    This respectable vision has a broad scope and does not define how it can be implemented.
    Thus, we observe the need for an alternative interaction paradigm with a more focused scope. Its vision integrates threads of similar ideas that currently exist as fragments in the human-computer interaction (HCI) space.
    
    We introduce Heads-Up Computing, a wearable platform-based interaction paradigm for which the ultimate goal of seamless and synergistic integration with everyday activities can be fulfilled.
    Heads-Up Computing \textbf{focuses only on the users’ immediate perceptual space}.
    At any given time, the space in which the human can perceive through his/her senses is what we refer to as the perceptual space.
    The specified \textit{form} \ie{} hardware and software of Heads-Up Computing, provides a solid foundation to guide future implementations, effectively putting humans at center stage.
\section{The Heads-Up Computing Paradigm}

Three main characteristics define Heads-Up Computing: 1) body-compatible hardware components, 2) multimodal voice and gesture interaction, and 3) resource-aware interaction model.
Its overarching goal is to offer a more seamless, just-in-time, and intelligent computing support for humans’ daily activities.

    \subsection{Body-compatible hardware components}
    To address the shortcomings of device-centered design, Heads-Up computing will distribute the input and output modules of the device to match the human input and output channels.
    Leveraging the fact that our head and hands are the two most important sensing \& actuating hubs, Heads-Up computing introduces a quintessential design that comprises two main components: the head-piece and the hand-piece.
    In particular, smart glasses and earphones will directly provide visual and audio output to the human eyes and ears.
    Likewise, a microphone will receive audio input from humans, while a hand-worn device (\ie{} ring, wristband) will be used to receive manual input.
    Note that while we advocate the importance of a hand-piece.
    Current smartwatches and smartphones are not designed within the principle of Heads-Up computing, as they require the user to adjust their head and hand position to interact with the device, thus are not synergistic enough with our daily activities.
    Our current implementation of a hand-piece consists of a tiny ring mouse can be worn on the index finger to serve as a small trackpad for controlling a cursor, as demonstrated by EYEditor~\cite{Ghosh2020EYEditor} and Focal~\cite{Focals2019}.
    It provides a relatively rich set of gestures to work with, which can provide manual input for smart glasses.
    While this is a base setup, many additional capabilities can be integrated into the smart glasses (\eg{} eye-tracking \cite{Mulvey2021}, emotion-sensing \cite{Hernandez2014}) and the ring mouse (\eg{} multi-finger gesture sensing, vibration output) for more advanced interactions.
    For individuals with limited body functionalities, Heads-Up computing can be custom designed to redistribute the input and output channels to the person's available input/output capabilities.
    For example, in the case of visually impaired individuals, the Heads-Up platform can focus on the capability of audio output with the earphone, and haptic input from the ring mouse to make it easier for them to access digital information in everyday living.
    Heads-Up computing exemplifies a potential design paradigm for the next generation, which necessitates a style of interaction highly compatible with natural body capabilities in diverse contexts.
    
    \subsection{Multimodal voice + gesture interaction}
    Similar to hardware components, every interaction paradigm also introduces new interactive approaches and interfaces.
    With a head- and hand-piece in place, users would have the option to input commands through various modalities: gaze, voice, and gestures of the head, mouth, and fingers.
    But given the technical limitations \cite{Lee2018Survey}, like being error-prone, requiring frequent calibration, and involving obtrusive hardware, it seems like only voice and finger gestures are currently the more promising modalities to be exploited for Heads-Up computing.
    As mentioned previously, voice modality is an underutilized input method that 1) is convenient, 2) allows us to free up both hands and vision, potentially, for doing another activity simultaneously, 3) is faster, and 4) is more intuitive.
    However, one of its drawbacks is that voice input can be inappropriate in noisy environments and sometimes socially awkward to perform \cite{Kollee2014}.
    Hence, it has become more important than ever to consider how users could exploit subtle gestures to employ less effort for input and generally do less.
    In fact, a recent preliminary study \cite{Sapkota2021} revealed that thumb-index-finger gestures could offer an overall balance and were preferred as a cross-scenario subtle interaction technique.
    More studies need to be conducted to maximize the synergy of finger gestures or other subtle interaction designs during everyday scenarios.
    For now, the complementary voice and gestural input method is a good starting point to support Heads-Up computing.
    It has been demonstrated by EYEditor~\cite{Ghosh2020EYEditor}, which facilitates on-the-go text-editing by using 1) voice to insert and modify text and 2) finger gestures to select and navigate the text.
    When compared to standard smartphone-based solutions, participants could correct text significantly faster while maintaining a higher average walking speed with EYEditor.
    Overall, we are optimistic about the applicability of multimodal voice + gesture interaction across many hand-busy scenarios and the general active lifestyle of humans.
    
    \subsection{Resource-aware interaction model}
    The final piece of Heads-Up Computing is its software framework. This framework allows the system to understand \textit{when} to use \textit{which} human resource.
    
    Firstly, the ability to sense and recognize ADL is made possible by applying deep learning approaches to audio \cite{Liang2019} and visual \cite{Nguyen2016} recordings.
    The head- and hand-piece configuration can be embedded with wearable sensors to infer the status of both the user and the environment.
    For instance, is the user stationary, walking, or running?
    What are the noise levels and lighting levels of the space occupied by the user?
    These are essential factors that could influence users’ ability to take in information.
    In the context of on-the-go video learning, \citet{Ram2021LSVP} recommended visual information be presented serially, persistently, and against a transparent background, to better distribute users’ attention between learning and walking tasks.
    But more can be done to investigate the effects of various mobility speeds \cite{Fennedy2020} on performance and preference of visual presentation style.
    It is also unclear how audio channels can be used to offload visual processing.
    Subtler forms of output like haptic feedback can also be used for low priority message notifications \cite{Roumen2015NotiRing}, or remain in the background of primary tasks \cite{Vaananen2014}.
    
    Secondly, the resource-aware system integrates \textit{feedforward} concepts \cite{Djajadiningrat2002} when communicating to users.
    It presents what commands are available and how they can be invoked.
    While designers may want to minimize visual clutter on the smart glasses, it is also important that relevant head- and hand-piece functions are made known to users.
    To manage this, the system needs to assess resource availability for each human input channel in any particular situation.
    For instance, to update a marathon runner about his/her physiological status, the system should sense if finger gestures or audio commands are optimal, and the front-end interface dynamically configures its interaction accordingly.
    Previous works primarily explored feedforward for finger/hand gestural input \cite{Jung2020, Fennedy2021}, but to the best of our knowledge, none have yet to address this growing need for voice input.\\
    
    An important area of expansion for the Heads-Up paradigm is its quantitative model, one that could optimize interactions with the system by predicting the relationship between human perceptual space constraints and primary tasks.
    Such a model will be responsible for delivering just-in-time information to and from the head- and hand-piece.
    We hope future developers can leverage essential back-end capabilities through this model as they write their applications.
    The resource-aware interaction model holds great potential for research expansion, and presents exciting opportunities for Heads-Up technology of the future.
\section{A Day in the Life with Heads-Up Computing}

Beth is a mother of two who works from home.
She starts her day by preparing breakfast for the family.
Today, she sets out to cook a new dish, the broccoli frittata.
With the help of a Heads-Up computing virtual assistant named Tom, Beth voices out, “Hey Tom, what are the ingredients for broccoli frittata?”
Tom renders an ingredient checklist on Beth's smart glasses.
Through the smart glasses' front camera, Tom sees what Beth sees, and detects that she is scanning the fridge.
This intelligent sensing prompts Tom to update the checklist collaboratively with Beth as she removes each ingredient from the fridge and places it on the countertop, occasionally glancing at her see-through display to double-check that each item matches.
With advanced computer vision and Augmented Reality (AR) capabilities, Beth can even request Tom to annotate where each ingredient is located within her sight.
Once all the ingredients have been identified, Beth proceeds with the cooking.
Hoping to be guided with step-by-step instructions, she speaks out: “Hey Tom, show me how to cook the ingredients.”
Tom searches for the relevant video on YouTube and automatically cuts it into step-wise segments, playing the audio through the wireless earset and video through the display.
Beth toggles the ring mouse she is wearing to jump forward or backward from the video.
Despite requiring both hands for cooking, she can utilize her idle thumb to control the playback of the video tutorial simultaneously.
Tom’s in-time assistance seamlessly adapts to Beth’s dynamic needs and constraints without referring to her remote phone, which would pause her task progress.

\begin{figure}
    \centering
	\includegraphics[width=0.9\textwidth]{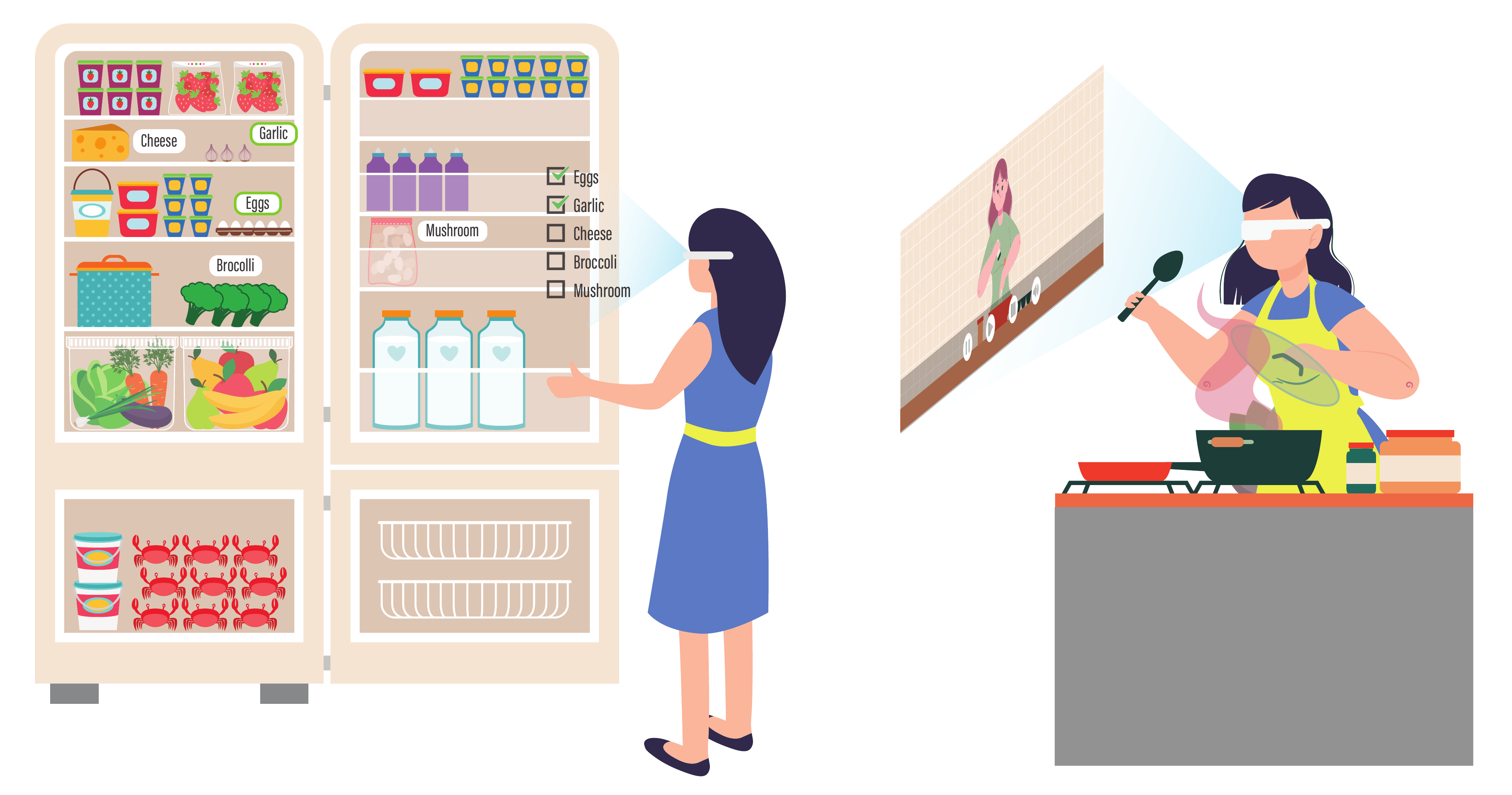}
	\caption{(Left) The user is browsing for ingredients within the fridge with the help of augmented labels. (Right) The user cooks the ingredients while simultaneously adjusting the playback of a guided video.}
	\Description{}
	\label{fig:aDayInTheLife} 
\end{figure}

Beth finishes with the cooking and feeds her kids, during which she receives an email from her work supervisor, asking for her available timing for an emergency meeting.
Based on Beth’s previous preferences, Tom understands that Beth values quality time with her family and prevents her from being bombarded by notifications from work or social groups during certain times of the day.
However, she makes an exception for messages labeled as `emergency.'
Like an intelligent observer, Tom adjusts information delivery to Beth by saying, “You have just received an emergency email from George Mason.
Would you like me to read it out?”.
Beth can easily vocalize “Yes” or “No” based on what suits her.
By leveraging the idle ears and mouth, Heads-Up computing allows Beth to focus her eyes and hands on what matters more in that context: her family.

Existing voice assistants such as Amazon Alexa, Google Assistant, Siri from Apple, and Samsung Bixby have gained worldwide popularity for the conversational interaction style that they offer.
They can be defined as “software agents that are powered by artificial intelligence and assist people with information searches, decision-making efforts or executing certain tasks using natural language in a spoken format” \cite{Ki2020}.
Despite allowing users to multitask and work hands-free, the usability of current speech-based systems still varies greatly \cite{Zwakman2021}.
These voice assistants currently do not achieve the depth of personalization and integration that Heads-Up computing can achieve given its narrower focus on the users’ immediate perceptual space and a clearly defined form \ie{} hardware and software.\\

The story above depicts a system that leverages visual, auditory, and movement-based data from a distributed range of sensors on the user’s body.
It adopts a first-person view as it collects and analyzes contextual information (\eg{} the camera on the glasses sees what the user sees and the microphone on the headpiece hears what the user hears).
It leverages the resource-aware interaction model to optimize the allocation of Beth’s bodily resources based on the constraints of her activities.
The relevance and richness of data collected from the user’s immediate environment, coupled with the processing capability, allows the system to anticipate and to calculate quantitative information ahead of the user.
Overall, we envision that Tom will be a human-like agent, able to interact with and assist humans.
From cooking to commuting, we believe that providing just-in-time assistance has the potential to transform relationships between devices and humans, thereby improving the way we live, learn, work, and play.
\section{Future of Heads-Up Computing}

Global technology giants like Meta, Google, and Microsoft have invested a considerable amount in the development of wearable AR \cite{applin_facebooks_2021}. The rising predicted market value of AR smart glasses \cite{Merel2017} highlights the potential of such interactive platforms.
As computational and human systems continue to advance, design, ethical, and socio-economical design challenges will also evolve.
We recommend the paper by \citet{Mueller2020}, which presents a vital set of challenges relating to Human-Computer Integration.
These include compatibility with humans, effects on human society, identity, and behavior. In addition, \citet{Lee2018Survey} effectively sum up interaction challenges specific to smart glasses and their implications on multi-modal input methods, all of which are relevant to the Heads-Up paradigm.

At the point of writing, there remains a great deal of uncertainty around wearable technology regulation globally.
Numerous countries have no regulatory framework, whereas existing frameworks in other countries are being actively refined \cite{bronneke_regulatory_2021}.
For the promise of wearable technology to be fully realized, we share the hope that different stakeholders — theorists, designers, and policymakers — collaborate to drive this vision forward and into a space of greater social acceptability.
As a summary of the Heads-Up Computing vision, we flesh out the following key points:

\begin{itemize}
  \item Heads-Up Computing is interested in moving away from device-centred computing and instead, placing humans at the centre stage. We envision a more synergistic integration of devices into the human's daily activities by first considering the human’s resource availability, primary and secondary task requirements, as well as resource allocation.
  \item Heads-Up Computing is a wearable platform-based interaction paradigm. Its quintessential body-compatible hardware components comprise a head- and hand-piece. In particular, smart glasses and earphones will directly provide visual and audio output to the human eyes and ears. Likewise, a microphone will receive audio input from humans, while a hand-worn device (i.e. ring, wristband) will be used to receive manual input.
  \item Heads-Up Computing utilizes multimodal I/O channels to facilitate multi-tasking scenarios. Voice input and thumb-index-finger gestures are examples of interactions that have been explored as part of the paradigm.
  \item The resource-aware interaction model is the software framework of Heads-Up Computing, which allows the system to understand when to use which human resource. Factors such as whether the user is in a noisy place can influence their ability to absorb information. Thus, the Heads-Up system aims to sense and recognize the user’s immediate perceptual space. Such a model will predict human perceptual space constraints and primary task engagement, then deliver just-in-time information to and from the head- and hand-piece.
  \item A highly seamless, just-in-time, and intelligent computing system has the potential to transform relationships between devices and humans, and there is a wide variety of daily scenarios for which the Heads-Up vision can translate and benefit. Its evolution is inevitably tied to  the development of head-mounted wearables, as this emergent platform makes its way into the mass consumer market.
\end{itemize}

When queried about the larger significance of the Heads-Up vision, the authors reflect on a regular weekday in their lives — 8 hours spent in front of a computer and another 2 hours on the smartphone.
Achievements in digital productivity come too often at the cost of being removed from the `real world'. 
What wonderful digital technology humans have come to create, perhaps the most significant in the history of our co-evolution with tools.
Could computing systems be so well-integrated that it not only supports but enhances our experience of physical reality?
The ability to straddle both worlds - the digital and non-digital one - is increasingly pertinent, and we believe it is time for a shift in paradigm.
We invite individuals and organisations to join us in our journey to design for more seamless computing support, improving the way future generations live, learn, work and play.


\bibliographystyle{ACM-Reference-Format}
\bibliography{references}


\end{document}